\documentclass[aps,nofootinbib,preprintnumbers,amsmath,amssymb]{revtex4}
\usepackage{epsfig}

\begin{document}
\preprint{USM-TH-168}
\title{Muonium-antimuonium conversion in
models with heavy neutrinos}
\author{Gorazd Cveti\v c}
  \email{gorazd.cvetic@usm.cl}
\author{Claudio O. Dib}
  \email{claudio.dib@usm.cl}
\affiliation{Dept.~of Physics, Universidad T\'ecnica
Federico Santa Mar\'{\i}a, Valpara\'{\i}so, Chile}
\author{C.~S.~Kim}
   \email{cskim@yonsei.ac.kr}
\affiliation{Department of Physics, Yonsei University,
Seoul 120-749, Korea}
\author{J.~D.~Kim}
    \email{jade@phya.snu.ac.kr}
\affiliation{Department of Physics, Seoul National University,
Seoul 151-742, Korea}

\date{11 April 2005}

\begin{abstract}
We study muonium-antimuonium conversion and $\mu^+e^-\to\mu^-e^+$
scattering within two different lepton-flavor-violating models
with heavy neutrinos: model I is a typical seesaw that violates
lepton number as well as flavor; model II has a neutrino mass
texture where lepton number is conserved. We look for the largest
possible amplitudes of these processes that are consistent with
current bounds. We find that model I has very limited chance of
providing an observable signal, except if a finely tuned condition
in parameter space occurs. Model II, on the other hand, requires
no fine tuning and could cause larger effects. However, the
maximum amplitude provided by this model is still two orders of
magnitude below the sensitivity of current experiments: one
predicts an effective coupling $G_{M\bar M}$ up to $10^{-4}G_F$
for heavy neutrino masses near 10 TeV. We have also clarified some
discrepancies in previous literature on this subject.
\end{abstract}

\maketitle

\section{Introduction}
\label{sec:intr}

Muonium is a hydrogenlike system of an electron ($e^-$) bound to a
positive muon ($\mu^+$) and antimuonium is the bound state of the
respective antiparticles ($\mu^- e^+$). It is formed as muons stop
inside matter, although the fraction of stopped muons that
actually form muonium depends very much on the material itself.
Since it has no hadronic constituents, muonium is an ideal place
to test electroweak interactions. In particular, since conversion
of muonium into antimuonium is completely forbidden within the
standard model (SM) because it violates lepton flavor, its
observation will be a clear signal of physics beyond the SM. This
phenomenon was first suggested almost 50 years ago
\cite{Pontecorvo} and since then several studies \cite{Feinberg,
Halprin} and experimental searches \cite{Willmann} have been done.
Many extensions of the SM could cause muonium to antimuonium
conversion \cite{Jungmann} (left-right symmetric gauge models,
extra neutrinos, extra higgses, SUSY, bileptonic gauge bosons,
etc).

Here we want to study the possibility that heavy Majorana
neutrinos in models of seesaw type could cause an observable
signal of muonium-antimuonium conversion. In previous works we
have studied the prospects for these models to generate tau decays
that violate lepton flavor \cite{CDKK}. The main feature of these
models is that they only extend the standard model by the
existence of extra neutral leptons. The most natural scenario is
then that lepton families will mix (thus destroying lepton flavor
conservation) and that neutrinos will acquire masses. In order to
insure that the standard neutrinos remain light, the texture of
the neutrino mass matrix should favor a seesaw type of mechanism.
In these models, muonium to antimuonium conversion appears at one
loop level in the perturbative expansion (box diagrams --see
Figs.\ 1 and 2) and it is thus a very small effect.

A simple estimate of this conversion probability within the models
in question shows that it is too small to be observed in
foreseeable experiments \cite{Ilakovac}. Another work along these
lines \cite{Clark}, making a rough estimate shows that present
experimental bounds set a lower bound on large Majorana masses at
around $10^1$ TeV. However, we find that the validity of the
calculation (perturbation theory to one loop) breaks down for
masses above that same scale. Therefore one can make clear
predictions for masses up to that scale only. We analyze the
process in more detail within the one-loop approximation,
performing a comprehensive search for the bounds in the currently
allowed parameter space. We confirm the conclusion that seesaw
models give a conversion probability that is still too small to be
at the reach of experiments to date. We also find that typical
seesaw models of heavy neutrinos are in most cases totally out of
reach of any foreseeable experiment, except in very special,
fine-tuned regions of the parameter space. In contrast, models
with heavy neutrinos that conserve lepton number give conversion
probabilities closer to present experimental sensitivities for
natural values of parameters.

In section II we give the generalities of muonium-antimuonium
conversion; in section III we calculate the process within the
models in question and do a comprehensive search of the presently
allowed parameter space, looking for the maximal possible
conversion probabilities. In section IV we give brief account of
the unbound collision $\mu^+ e^-\to \mu^- e^+$ and its prospects.
In section V we state our conclusions. Details of the seesaw
models and loop calculations are included as appendices.

\section{Muonium to antimuonium Conversion}
\label{MbM}

Muonium $|M \rangle$ is a nonrelativistic Coulombic bound state of
$\mu^+$ and $e^-$, and antimuonium $|{\overline M} \rangle$ a
similar bound state of $\mu^-$ and $e^+$. The nontrivial mixing
between $ |M \rangle$ and $|{\overline M} \rangle$ comes in our
context from the non-vanishing LFV transition amplitude for $e^-
\mu^+ \to e^+ \mu^-$. In a model with heavy neutrinos, this
amplitude appears at one loop (box diagrams) in the electroweak
interaction extended with extra (heavy) neutrinos. This transition
is usually expressed in terms of a local effective Hamiltonian
density with an effective coupling $G_{M\bar M}$:
\begin{equation}
{\cal H}_{\rm eff}(x) = \frac{G_{ {\overline M} M}}{ \sqrt{2} }
\left[ {\overline \mu}(x) \gamma^{\alpha} (1\!-\!\gamma_5) e(x)
\right] \left[ {\overline \mu}(x) \gamma_{\alpha} (1\!-\!\gamma_5)
e(x) \right] \ .
\label{Leff}
\end{equation}

Due to this interaction, $|M\rangle$ and ${\overline M}\rangle$
are not mass eigenstates. In this basis, the mass matrix has
non-diagonal components:
\begin{equation}
m_{M\bar M} = \langle M | \int d^3 x\ {\cal H}({\bf x})|\bar
M\rangle ,
\end{equation}
where the (non-relativistic) bound states are chosen with unit
norm in a volume $V$, namely $\langle M({\bf P})|M({\bf
P'})\rangle =$ $(2\pi)^3 \delta^3({\bf P}-{\bf P'})/V$. If we
express the bound state in terms of its constituents with a
momentum distribution $f(p)$:
\begin{equation}
|M(0)\rangle = \int \frac{d^3 p}{(2\pi)^3} f(p) a_p^{(e)\dagger}
b_{-p}^{(\mu)\dagger}|0\rangle ,
\end{equation}
and neglect the momentum dependence in the spinors, the mass
mixing element is:
\begin{equation}
m_{M\bar M} = 16 \frac{G_{M\bar M}}{\sqrt{2}}\left|\int \frac{d^3
p}{(2\pi)^3} f(p)\right|^2 \bar u_{(\mu)L} \gamma^\alpha u_{(e)L}
\bar v_{(\mu)L} \gamma_\alpha v_{(e)L}
\end{equation}
Since angular momentum is conserved, the only non vanishing
elements are those where the spin of the initial and final bound
state are the same. For both the singlet or each of the triplet
components, the spinor product has the same value $2m_\mu m_e$
\cite{spinor}. Additionally, the integral of $f(p)$ is the spatial
wavefunction at zero distance, $|\int d^3 p/(2\pi)^3 f(p)|^2 =
|\Psi(0)|^2/(2m_\mu 2 m_e)$. The mass mixing element
\cite{Fein-Wein} thus reduces to
\begin{equation}
m_{M\bar M} = 16 \frac{G_{M\bar
M}}{\sqrt{2}}\frac{\left|\Psi(0)\right|^2}{2}\quad = 16
\frac{G_{M\bar
M}}{\sqrt{2}}\frac{(\mu\alpha)^3}{2\pi}
\label{Mixing}
\end{equation}
where $\mu \approx m_e$ is the muonium reduced mass and $\alpha$
the fine structure constant. In vacuum, the diagonal mass terms
are equal, $m_{MM} = m_{\bar M \bar  M}$. The mass eigenstates are
then the simple combinations $|M_1\rangle = (|M\rangle + |\bar
M\rangle)/\sqrt{2}$ and $|M_2\rangle = (|M\rangle - |\bar
M\rangle)/\sqrt{2}$ and the mass eigenvalues differ in $\Delta
m\equiv$ $m_1 - m_2 = 2\ m_{M\bar M}$. This is the typical case of
a quantum system of two states that oscillate into each other
under time evolution, like the $K^0-\bar K^0$ system in hadrons.
However, unlike the latter, in practical terms muonia do not have
time to show an oscillation, because their oscillation time is
much longer than their decay time. So one can only hope to observe
the mixing phenomenon by measuring the probability that a state
that starts as a Muonium ($\mu^+ e^-$) decays as an Antimuonium
(decay into a high energy electron and low energy positron). So,
the state that starts as $|M\rangle$ at $t=0$ (call it
$|M(t)\rangle$) evolves as:
\begin{equation}
|M(t)\rangle = |M_1\rangle \langle M_1|M\rangle e^{-i\ m_1 t} +
|M_2\rangle \langle M_2|M\rangle e^{-i\ m_2 t}
\end{equation}
and will decay as a $\bar M$ with a total probability:
\begin{equation}
P(M\to \bar M) = \int_0 ^\infty \frac{dt}{\tau} e^{-t/\tau}
|\langle \bar M|M(t)\rangle|^2 \quad = \frac{(\Delta m\
\tau)^2}{2(1+(\Delta m\ \tau)^2)} \approx \frac{1}{2} (\Delta m\
\tau)^2
\label{prob-delta}
\end{equation}
where $\tau$ is the muon lifetime. Eq.\ (\ref{prob-delta}) does
not take into account the effect of static electromagnetic fields
in materials, which break the degeneracy $m_{MM} = m_{\bar M\bar
M}$ and cause an extra suppression on the
probability\cite{Feinberg}, an important effect in some
experiments. In any case, we see from Eq.\ (\ref{prob-delta}) that
the conversion probability is in general very small and
proportional to $|G_{M\bar M}|^2$:
\begin{equation}
P(M\to \bar M) = \frac{64\ G_F^2\ \alpha^6 m_e^6\ \tau^2}{\pi^2}
\left(\frac{G_{M\bar M}}{G_F}\right)^2\quad =2.64\times 10^{-5}
\left(\frac{G_{M\bar M}}{G_F}\right)^2.
\label{probability}
\end{equation}

\begin{table}
  \begin{tabular}{|l l l|}  \hline
  $P(M\to \bar M)$ & \quad$G_{M\bar M}/G_F$ &\quad Experiment \\ \hline
  $< 2.1\times 10^{-6}$ & \quad$< 0.29$ & \quad Huber et al. (1990)\\
  $< 6.5\times 10^{-7}$ & \quad$< 0.16$ & \quad Matthias et al. (1991)\\
  $< 8.0\times 10^{-9}$ & \quad$< 1.8\times 10^{-2}$ & \quad
  Abela et al. (1996)\\
  $< 8.3\times 10^{-11}$ & \quad $< 3\times 10^{-3}$ & \quad Willmann
et al. (1999)\\ \hline
  \end{tabular}
  \caption{Experimental results (see Ref. \cite{Willmann})
           for the conversion probability and the deduced
           upper bound for the coupling $G_{M\bar M}$ (this bound on the
           last row is weaker than what Eq.\ (\ref{probability}) gives, because of
           magnetic field suppression).}
  \end{table}

The experiments so far have gradually been
 improving on an upper bound for $P(M\to \bar M)$ and thus for
$G_{M\bar M}$ (See Table 1), but no positive signal of this lepton
flavor violating process has ever been observed, which is
consistent with the null prediction of the standard model. All new
physics in this process enters in the parameter $G_{M\bar M}$. We
will now examine the prospects for $G_{M\bar M}$ in extensions of
the standard model with extra heavy neutrinos originating in
seesaw mechanisms.

\section{Prediction of the Seesaw models}
\label{sec:pred}

We want to explore the prediction of lepton flavor violating
seesaw models for muonium-antimuonium conversion. The details of
the models are given in Appendix A. The main features of these
models are that (i) they contain heavy neutrinos, (ii) in their
mass basis, the heavy neutrinos couple to charged leptons of
different generations under weak interactions (just like the
Cabbibo-Kobayashi-Maskawa prescription in the quark sector), thus
inducing flavor changing processes (iii) in general the neutrino
masses are of Majorana type.

The mixing matrix $B_{li}$, which is defined in Appendix A, is the
coefficient of the left-handed current that connects a charged
lepton of flavor $l = e,\ \mu,\ldots$ with a neutrino of flavor
$i=1,\ 2,\ldots$\ \ In a model with $N_L$ charged leptons, the
first $N_L$ neutrinos are light (standard) and the rest are heavy.
Here we will only consider $N_L=2$ to make the estimates. Of
course, general lepton flavor mixing would induce many kinds of
processes that are not observed in the real world. The
non-observation of those processes impose severe bounds on the
values of $B_{l i}$. Here we use the results of the analysis of
Ref.\ (\cite{Nardi}), which establish bounds on combinations of
mixing elements characterized by
$(s^{\nu_l}_L)^2\equiv\sum_{i=h}|B_{li}|^2$, where $h$ indicates
heavy neutrinos, singlets under $SU(2)_L$:
\begin{equation}
(s^{\nu_e}_L)^2 < 0.005,\ \ \ (s^{\nu_\mu}_L)^2 <0.002,\ \ \
(s^{\nu_\tau}_L)^2 < 0.010 .
\label{s2s}
\end{equation}

As we said, in these models, muonium-antimuonium conversion is
induced by the process $\mu^+ e^-\to \mu^- e^+$ at one-loop via
box diagrams (see Figs.\ 1 and 2), where the neutrinos appear as
internal quanta. The effective coupling $G_{M\bar M}$ is a
function of the internal neutrino masses $m_i$ and mixing
parameters $B_{li}$ that arises from the integrals of the box
diagrams:
\begin{eqnarray}
{G_{M\bar M}}/{G_F} =  \frac{\alpha_W}{32\pi}  \sum_{i,j = all\
\nu's} \Big[ 2 B^*_{e i} B^*_{e j} B_{\mu i} B_{\mu j}\ F_{\rm
Box}(x_i, x_j) - (B^*_{e i})^2 (B_{\mu j})^2\lambda_i^* \lambda_j\
G_{\rm Box}(x_i,x_j) \Big] \ , \label{FBoxeuue}
\end{eqnarray}
where $\alpha_W = g^2/4\pi$ is the weak $SU(2)_L$ fine structure
constant,  $x_i = m_i^2/M_W^2$ are the squares of neutrino masses
inside the loop in units of the $W$ boson mass, $\lambda_i$ are
``creation phases'' that appear in Majorana fields \cite{Kayserb},
and $F_{\rm Box}(x_i, x_j)$ and $G_{\rm Box}(x_i, x_j)$ are the
loop functions that appear integrating the ``Dirac boxes" of
Fig.~1 and the ``Majorana boxes" of Fig.~2, respectively (external
masses and momenta inside the integrals are neglected as compared
to loop momenta).  These functions, symmetric under the
interchange $x_i \leftrightarrow x_j$, are:
\begin{eqnarray}
F_{\rm Box}(x_i, x_j) &=& \frac{1}{x_i - x_j} \bigg\{
\left(1+\frac{x_i
x_j}{4}\right)\left(\frac{1}{1-x_i}+\frac{x_i^2}{(1-x_i)^2}\ln
x_i\right) - 2x_i
x_j\left(\frac{1}{1-x_i}+\frac{x_i}{(1-x_i)^2}\ln
x_i\right)\nonumber\\&& \qquad\qquad\quad -(i\to
j)\bigg\}
\label{fbox}
\end{eqnarray}
\begin{eqnarray}
G_{\rm Box}(x_i, x_j) &=& -\frac{\sqrt{x_i x_j}}{x_i - x_j}
\bigg\{ \left(4+ x_i x_j \right)
\left(\frac{1}{1-x_i}+\frac{x_i}{(1-x_i)^2}\ln x_i\right) - 2
\left(\frac{1}{1-x_i}+\frac{x_i^2}{(1-x_i)^2}\ln
x_i\right)\nonumber\\&&\qquad\qquad\quad -(i\to
j)\bigg\}
\label{gbox}
\end{eqnarray}
\begin{center}
\begin{figure}[htb]
\epsfig{file=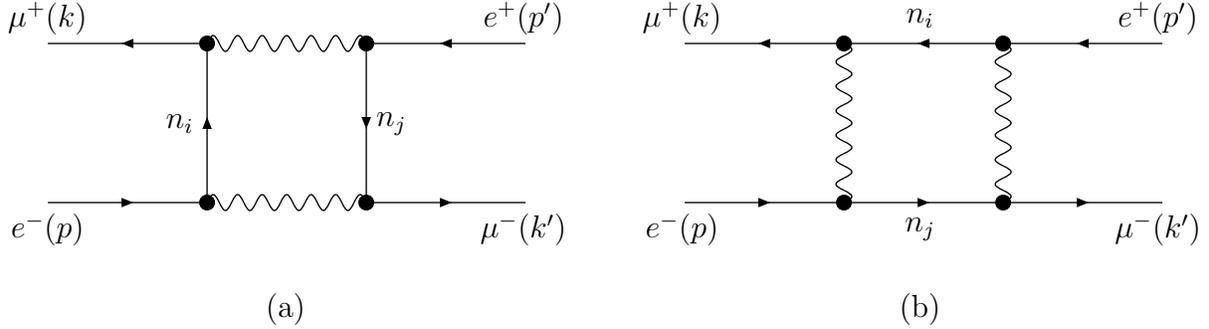}
 \caption{ ``Dirac boxes": box diagrams for generic (Dirac or Majorana) neutrinos
 (each wavy line is either a $W$ or a charged Goldstone boson if we work in the Feynman gauge). }
 \label{fig1}
\end{figure}
\end{center}
\begin{center}
\begin{figure}[htb]
\epsfig{file=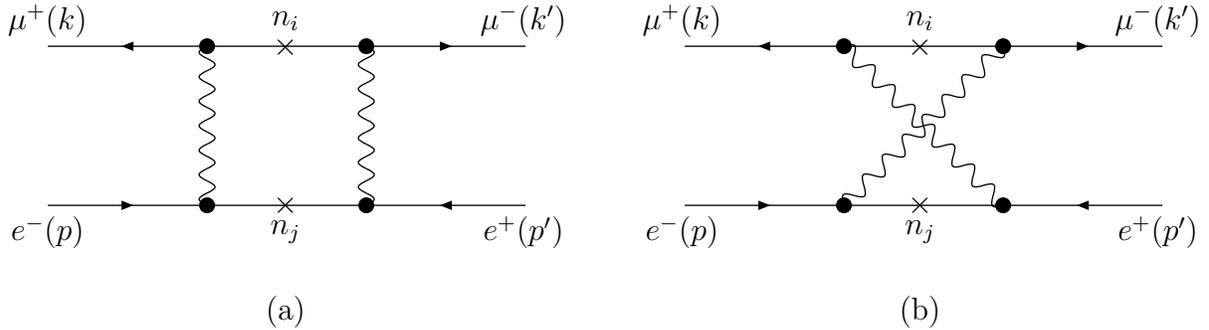}
 \caption{ ``Majorana boxes": box diagrams for Majorana neutrinos only (each
wavy line is either a $W$ or a charged Goldstone boson). }
\label{fig2}
\end{figure}
\end{center}
We should point out that we have a discrepancy in the global sign
in front of $G_{\rm Box}$ in Eq.\ (\ref{FBoxeuue}) compared to
Ref.\ \cite{IP}. It is important to remark that this global sign
can not be changed by simple redefinition of fermion phases. For
example, one can redefine the phases of the Majorana fields and
choose $\lambda_i = +1$ for all $i$, or equivalently one can
absorb those phases in the mixing matrix $B_{li}$, but it is
impossible to choose the phases so that $\lambda_i^* \lambda_j =
-1$ for all $i,j$. It is also important to notice a discrepancy
with Ref.\ (\cite{Clark}), where the authors claim that the
contribution from the two diagrams of Fig.~2 with Majorana
neutrinos ({\it i.e.} the $G_{\rm Box}$ term) cancel because of
Fermi statistics. Actually, when one carries through the
calculation, the minus sign from Fermi statistics is compensated
by the inner structure of the diagrams and at the end both
diagrams are equal and they add! Since $G_{\rm Box}$ dominates
over $F_{\rm Box}$ for large neutrino masses, this could be an
important issue in the estimates. To clarify these matters, we
include the derivation of Eq.\ (\ref{FBoxeuue}) in Appendix B.

Using the unitarity of the mixing matrix, Eq.~(\ref{FBoxeuue}) can
be rewritten as a double sum over heavy neutrino flavors only
\begin{eqnarray}
{G_{M\bar M}}/{G_F} &=&  \frac{\alpha_W}{32\pi}  \sum_{I,J =
N_L+1}^{N_L+N_H} \Big\{ 2 B^*_{e I} B^*_{e J} B_{\mu I} B_{\mu J}
\ \left[ F_{\rm Box}(x_I, x_J) - 2 F_{\rm Box}(x_I,0) + F_{\rm
Box}(0,0) \right] \nonumber\\ &&- (B^*_{e I})^2 (B_{\mu
J})^2\lambda_I^* \lambda_J\ G_{\rm Box}(x_I,x_J) \Big\} \ ,
\label{FBoxred}
\end{eqnarray}
where light neutrino masses are neglected.

Table I shows that in order for muonium to antimuonium conversion
to be observable in at least the most sensitive experiment to
date, the coupling ${G_{M\bar M}}/{G_F}$ should be at least ${\cal
O}(10^{-3})$. A rough estimate for the largest value expected for
$G_{M\bar M}$ from seesaw models of heavy neutrinos, and
consistent with the present bounds of Eq.\ (\ref{s2s}), can be
obtained considering that the functions $F_{\rm Box}$ and $G_{\rm
Box}$ of Eqs.\ (\ref{fbox}) and (\ref{gbox}) are growing functions
for large neutrino masses and that the mixings in seesaw models
are in general decreasing functions of the neutrino masses. For
neutrino masses of a few TeV, $F_{\rm Box}$ and $G_{\rm Box}$ are
of order $10^1$ and $10^3$ respectively and we will assume that
for those masses the mixings can still be near the maxima shown in
Eq.\ (\ref{s2s}). In that case the process is dominated by $G_{\rm
Box}$ and we obtain
\begin{equation}
{G_{M\bar M}}/{G_F}\ \sim \  \frac{\alpha_W}{32\pi}  G_{\rm
Box}\times  (s^{\nu_\mu}_L)^2
 (s^{\nu_e}_L)^2 \quad \sim \  10^{-6}
\label{rough}
\end{equation}
which is 3 orders of magnitude smaller than present observable
values.

If we want to search for larger possible values for ${G_{M\bar
M}}$, we should assume larger masses for the heavy neutrinos.
However, the assumption of larger masses may enter into conflict
with the choice of mixing parameters at their current upper bounds
[Eq.\ (\ref{s2s})]. As mentioned earlier, in seesaw models the
mixings decrease as the neutrino masses are taken to be larger. If
the neutrino masses are very large, the mixings will be forced to
be considerably smaller than their current upper bound, and so
assuming larger masses may not help to get potentially larger
conversion probabilities. This effect was not an issue in Ref.\
\cite{Clark} because they assumed rather large Dirac lepton masses
($m_D \sim M_W$). With this assumption one naturally predicts
rather large conversion probabilities, however enters into
conflict somewhere else: the prediction for the light neutrino
masses, $m_\nu \sim m_D^2/M_h \sim 10^2$ MeV, now turns out to be
too large. Therefore, one needs to refine these estimates to be
consistent with all present bounds. Here we do it by finding the
relation between the neutrino masses and the mixings, following
the whole process of mass matrix diagonalization from the start
within each model. Keeping these consistency conditions, we search
for the largest possible values for ${G_{M\bar M}}$ in the
scenarios of models I and II.

\subsection{Maximal conversion amplitude in Model I}

The key problem is to find the neutrino mass matrix ${\cal M}$ in
Eq.~(\ref{Lnumass}) and the mixing parameters $B_{l i}$ in
Eq.~(\ref{B}), constrained by Eq.~(\ref{s2s}), that maximize the
effective coupling ${G_{M\bar M}}$.

The general form of ${\cal M}$ for two families in Model~I can be
described as:
\begin{equation}
{\cal M} = \left(
\begin{array}{cc}
0 & m_D \\ m_D^T & m_M
\end{array}
\right), \quad m_D= \left(
\begin{array}{cc}
a & b \; e^{i \delta_1}  \\ c \; e^{i \delta_2}&  d
\end{array}
\right), \quad m_M= \left(
\begin{array}{cc}
M_1 & 0 \\ 0 & M_2
\end{array}
\right)  \ , \label{Dm1}
\end{equation}
where $a$, $b$, $c$, $d$ are real Dirac masses, $\delta_j$
($j=1,2$) are two independent physical phases, and $M_1$ and $M_2$
(with $M_1<M_2$) are large masses. The mixing parameters $B_{l i}$
are related to the transformation matrix $U$ of
Eq.~(\ref{congruent}), which can be written approximately as a
product of a seesaw block matrix $U_s$ and a light-sector mixing
matrix $V^{\dagger}$
\begin{eqnarray}
U & = & V^{\dagger} U_s \ , \qquad U_s = \left[
\begin{array}{cc}
(I_{2 \times 2} - \frac{1}{2} m_D m_M^{-2} m_D^{\dagger}) & - m_D
m_M^{-1}
\\
m_M^{-1} m_D^{\dagger} & (I_{2 \times 2} - \frac{1}{2} m_M^{-1}
m_D^{T} m_D m_M^{-1} )
\end{array}
\right] \ , \label{Um1}
\\
V^{\dagger} &=& \left[
\begin{array}{cc}
V_{l}^{\dagger} & 0
\\
0 & I_{2 \times 2}
\end{array}
\right] \ , \qquad V_l^{\dagger} = \left(
\begin{array}{cc}
\cos \theta & \sin \theta \ e^{- i \varepsilon}
\\
- \sin \theta \ e^{i \varepsilon} & \cos \theta
\end{array}
\right) \ , \label{UsV}
\end{eqnarray}
where $\varepsilon$ is a CP phase. The transformation
(\ref{congruent}) with $U_s$ alone gives an approximately
block-diagonal mass matrix if we neglect the correction terms
${\cal O}(m_D^2/m_M^2)$ and ${\cal O}(m_D^3/m_M^2)$:
\begin{equation}
U_s {\cal M} U_s^T = \left[
\begin{array}{cc}
- m_D m_M^{-1} m_D^T \left( 1 + {\cal O}(m_D^2/m_M^2) \right) &
{\cal O}(m_D^3/m_M^2) \\ {\cal O}(m_D^3/m_M^2) & m_M \left( 1 +
{\cal O}(m_D^2/m_M^2) \right)
\end{array}
\right] \ . \label{blockm2}
\end{equation}
 The matrix $V_l^{\dagger}$, Eq.~(\ref{UsV}),
then generates a congruent diagonalization of the light-sector
matrix $m_D m_M^{-1} m_D^T$. In $V_l^{\dagger}$, $\theta=0$ and
$\pi/4$ correspond to zero and maximal $\nu_e$-$\nu_{\mu}$ mixing.
According to solar neutrino experiments, we have maximal
$\nu_e$-$\nu_{\mu}$ mixing, $\theta=\pi/4$. If we demand that this
maximal mixing be obtained independently of the large masses $M_1$
and $M_2$, then straightforward algebra gives the following simple
restrictions on the light Dirac sector parameters:
\begin{equation}
a^2 = c^2, \quad b^2 = d^2, \quad \delta_1 = \delta_2 \ (\equiv
\delta) \ . \label{maxmixm1}
\end{equation}
The value of $\delta$ can be restricted to $- \pi/2 < \delta \leq
\pi/2$. The  two light neutrino eigenmasses  are then
\begin{equation}
m_{\nu_e, \nu_{\mu}} = {\Bigg |} \left[ \left( \frac{a^2}{M_1}
\right)^2 \! + \! \left( \frac{b^2}{M_2} \right)^2  \! + \! 2
\frac{a^2}{M_1}\frac{b^2}{M_2} \cos(2 \delta) \right]^{1/2} \pm
\left( \frac{ac}{M_1} \! + \! \frac{bd}{M_2} \right) {\Bigg |} \ ,
\label{eigmm1}
\end{equation}
and the heavy-to-light mixing parameters $(s_L^{\nu_l})^2 \equiv
\sum_{h=3}^4 |B_{l h}|^2$ and $B_{l h}^{\ast} = U_{h l}$ are
\begin{eqnarray}
(s_L^{\nu_e})^2 & = & (s_L^{\nu_{\mu}})^2 = \frac{a^2}{M_1^2} +
\frac{b^2}{M_2^2} \; (\equiv s_L^2 < 0.002) \ , \label{sL2m1}
\\
B_{l h}^{\ast} & = & \left[
\begin{array}{ll}
a/M_1 \ , & (b/M_2) e^{- i \delta} \
\\
(c/M_1) e^{- i \delta} \ , & d/M_2 \
\end{array}
\right] \quad \ell = e,\, \mu\, ;\  h= 3,\, 4\label{Blh}
\end{eqnarray}
The elements of the diagonal matrix $\Lambda$ of
Eq.~(\ref{congruent}) generally are complex phase factors; here we
have $\lambda_3 = \lambda_4 = +1$ while $\lambda_1, \lambda_2
\not= +1$.

In order to reduce the parameter space, for every given $M_1$ and
$M_2$ we choose the values of the Dirac sector parameters ($a$,
$b$, $c$, $d$, $\delta_1$, $\delta_2$) that maximize the effective
coupling $|{G_{M\bar M}}|$ [Eq.~(\ref{FBoxred})], and then explore
the $M_1, M_2$ parameter space on top of that condition. On the
basis of Eq.~(\ref{FBoxred}), we can set an upper bound on
$|{G_{M\bar M}}|$, for given $M_1$ and $M_2$:
\begin{eqnarray}
 G_{M\bar M} / G_F  &\leq& \frac{\alpha_W}{32\pi}\left\{ \hat F_{\rm Box}
\times \left( \sum_{I=3}^4 |B^*_{e I} B_{\mu I} | \right)^2 + \hat
G_{\rm Box} \times \sum_{I=3}^4 |B^*_{e I}|^2 \sum_{J=3}^4 |B_{\mu
J} |^2 \right\}\label{FBoxleq1m1}
\\
&\leq& \frac{\alpha_W}{32\pi} ( \hat F_{\rm Box} + \hat G_{\rm
Box} ) \times
 \sum_{I=3}^4 |B^*_{e I}|^2  \sum_{J=3}^4 |B_{\mu J}|^2
 \nonumber\\
&\leq& \frac{\alpha_W}{32\pi} ( \hat F_{\rm Box} + \hat G_{\rm
Box} )  \times (s_L^{\nu_e})^2 (s_L^{\nu_{\mu}})^2 \ ,
\label{FBoxleq2m1}
\end{eqnarray}
where $\hat F_{\rm Box} = 2 \; {\rm max} |F_{\rm Box}(x_I, x_J)- 2
F_{\rm Box}(x_I,0) + F_{\rm Box}(0,0)|$, and $\hat G_{\rm Box} =
{\rm max} | G_{\rm Box}(x_I,x_J)|$. The maximum of ${G_{M\bar M}}$
is reached as it approaches this upper bound and simultaneously
the maximal values $(s_L^{\nu_e})^2 = (s_L^{\nu_{\mu}})^2 =
(s_L)^2_{\rm max} = 0.002$, Eqs.~(\ref{s2s}) and (\ref{sL2m1}),
are reached.

Now, there are two distinct cases into which the maximum mixing
condition (\ref{maxmixm1}) can be divided:

\begin{itemize}
\item
{\bf Case (a):} $c = \pm a$ and $d = \pm b$ (and $\delta_1 =
\delta_2 \equiv \delta$). In this case, $m_{\nu_{\mu}} \geq
(a^2/M_1 + b^2/M_2)$ by Eq.~(\ref{eigmm1}), and by
Eq.~(\ref{sL2m1}) consequently $s_L^2 \leq m_{\nu_{\mu}}/M_1 \leq
3 \cdot 10^{-11}$, where the latter inequality was obtained by
taking $M_1 \geq 100$ GeV and $m_{\nu_{\mu}}<3$ eV as suggested by
present neutrino experiments ($m_{\nu_e}<3$ eV,
Ref.~\cite{Eidelman:2004wy}; $\Delta m^2_{sol} = |m^2_{\nu_{\mu}}
- m^2_{\nu_e}| \sim 10^{-10} \ {\rm eV}^2 \approx 0$,
Refs.~\cite{bahcall}). This value of $s_L^2$ extremely suppresses
 ${G_{M\bar M}}$, as seen by Eq.~(\ref{FBoxleq2m1}),
so we will disregard this case for our purposes.
\item
{\bf Case (b):} $c = \pm a$ and $d = \mp b$ (and $\delta_1 =
\delta_2 \equiv \delta$). In this case, by Eq.~(\ref{eigmm1})
\begin{equation}
m_{\nu_e, \nu_{\mu}} = {\Bigg |} \left[ \left( \frac{a^2}{M_1}
\right)^2 \! + \! \left( \frac{b^2}{M_2} \right)^2  \! + \! 2
\frac{a^2}{M_1}\frac{b^2}{M_2} \cos(2 \delta) \right]^{1/2} \pm
{\bigg |} \frac{a^2}{M_1} \! - \! \frac{b^2}{M_2} {\bigg |} {\Bigg
|} \ . \label{eigm2m1}
\end{equation}
Therefore, $\Delta m^2_{sol} = m^2_{\nu_{\mu}} - m^2_{\nu_e} \geq
4 | a^2/M_1 - b^2/M_2 |^2$, i.e., $| a^2/M_1 - b^2/M_2 | \leq
(1/2) (\Delta m^2_{sol})^{1/2} \sim 10^{-5} \ {\rm eV}$, which is
very small compared to $m_e, m_\mu$. This means that solar
neutrino experiments imply an almost exact relation:
 \begin{eqnarray}
 a^2/M_1 = b^2/M_2, \label{rel2}
 \end{eqnarray}
 which in turn implies the following expression for the $\nu_\mu$ mass:
 \begin{eqnarray}
 m_{\nu_{\mu}}/M_1 = (a^2/M_1^2)
\sqrt{ 2 (1 + \cos 2 \delta )}. \label{numumass}
\end{eqnarray}
This is a very small quantity, of order $10^{-11}$. Here we see
the typical behavior of seesaw models: since $m_\nu/M_1 \sim
a^2/M_1^2$ and the mixing of heavy-to-light sectors is $B_{\ell
h}\sim a/M_1$, the latter vanishes as the disparity between light
vs. heavy neutrino masses intensifies, namely $m_\nu \ll M_1$.

However, we do not want $a/M_1$ to be too small, because it would
suppress the mixing parameter $s_L^2$ [see Eq.\, (\ref{sL2m1})].
Therefore, in view of Eq.\, (\ref{numumass}) we set $\delta\to
\pi/2$, which keeps the mass disparity without the vanishing of
mixing. This is a rather finely tuned condition, but we admit it
because we are only looking for maximal transition probabilities.
Then, using (\ref{sL2m1}), (\ref{rel2}) and the defining condition
of this case, namely $c=a$ and $d=-b$, we have set all the
parameters of the Dirac sector for maximal conversion probability
in terms of just $M_1$ and $M_2$:
\begin{equation}
a = c = M_1 (s_L)_{\rm max}/\sqrt{1 + M_1/M_2} \ , \quad b = - d =
a \sqrt{M_2/M_1}\ , \quad  \delta\to\pi/2 . \label{abmodI}
\end{equation}
\end{itemize}

We then proceed to scan the values of $|{G_{M\bar M}}|$ as a
function of $M_1$ and $M_2$ restricted to the conditions of Eq.\,
(\ref{abmodI}). We find, interesting enough, that the Majorana
boxes, {\it i.e.} terms proportional to $G_{\rm Box}(x_I, x_J)$,
do not dominate the amplitude for large masses --at least not
under condition (\ref{abmodI})-- even though the functions $G_{\rm
Box}(x_I, x_J)$ are individually larger than $F_{\rm Box}(x_I,
x_J)$. Instead, the mixing elements $B_{\ell h}$ that multiply
these functions are such that both $F_{\rm Box}$ and $G_{\rm Box}$
terms are of the same order and interfere. Indeed, if one expands
the functions $F_{\rm Box}$ and $G_{\rm Box}$ in the amplitude
(\ref{FBoxred}) for asymptotic values of $x_I$ and $x_J$, the
leading terms come na\"\i vely from $G_{\rm Box}$; however, these
are multiplied by a factor proportional to $(m_{\nu_{\mu}}/M_1)^2$
arising from the mixing elements, which totally suppress them,
leaving the leading terms of $F_{\rm Box}$ and the subleading
terms of $G_{\rm Box}$ at the same level of interference.

Fig.~3 shows the graphs for the maximal coupling $|{G_{M\bar M}}|$
as a function of $M_1$ and $M_2$, which is consistent with all
present constraints within model I.

\begin{center}
\begin{figure}[htb]
\epsfig{file=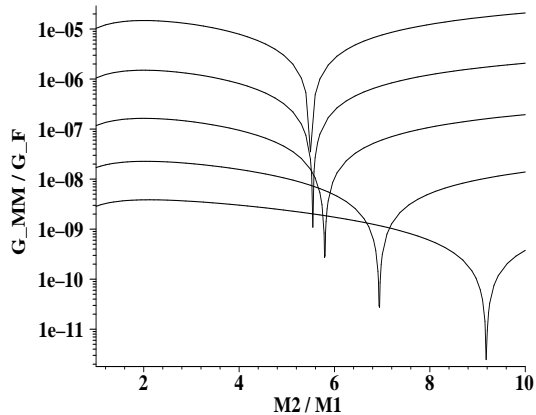,height=7.5 cm,width=6 cm,angle=-90}
\caption{ Maximal effective coupling $G_{M\bar M}/G_F$ for muonium
to antimuonium  conversion in Model~I, as a function of the ratio
of Majorana masses $M_2/M_1$, for different masses $M_1$: from the
lower to the upper curve, $M_1 =$ 100 GeV, 300 GeV, 1 TeV, 3 TeV,
10 TeV. The dips reflect an accidental cancellation of the leading
terms of Dirac and Majorana box diagrams for particular values of
$M_1$ and $M_2$. }
 \label{fig3}
\end{figure}
\end{center}

We see that the amplitude is indeed a growing function of $M_1,
M_2$ (we also see that for given $M_1$, local maxima are reached
if $M_2 \approx 2 M_1$). We do not push $M_1, M_2$ higher than
around 10 TeV, because the perturbative expansion to one loop
breaks down above that scale, as the authors of Ref.\, \cite{IP}
point out. They stated that the perturbative expansion is valid
provided $M_h^2 \sum_\ell |B_{\ell h}|^2 < 2 M_W^2 / \alpha_W$. In
our case, this condition becomes:
 \begin{eqnarray}
 M_1 M_2 < \frac{M_W^2}{\alpha_W s_L^2} \sim (10\ {\rm TeV})^2.
 \end{eqnarray}.

From Fig.~3, we see that the maximum expected value of $G_{M\bar
M}/G_F$ is around $10^{-5}\to 10{_4}$, which is still almost two
orders of magnitude below the level of sensitivity of current
experiments. To close the analysis, we may comment on the finely
tuned parameter $\delta\to \pi/2$. It is clear that if we do not
impose this condition, then the mixing elements $s_L^2$ are $\sim
a^2/M_1^2 \sim m_\nu/M_1$. For $m_{\nu_\mu}\sim 190$ KeV, which is
the upper bound \cite{Eidelman:2004wy}, and $M_1\sim 10$ TeV, we
get $s_L^2\sim 10^{-8}$, which signifies a suppression of 6 orders
of magnitude in the amplitude  with respect to the maxima found in
our analysis.

\subsection{Maximal conversion amplitude in Model II}

In Model II, Eq.~(\ref{M-model2}), with CP-conserving mass terms,
the submatrices are real and can be taken in the form
\begin{equation}
{\cal M} = \left(
\begin{array}{ccc}
0 & m_D &0 \\ m_D^T & 0 & m_M^T \\ 0 & m_M & 0
\end{array}
\right), \quad m_D= \left(
\begin{array}{cc}
a & b  \\ c &  d
\end{array}
\right), \;\;\; m_M= \left(
\begin{array}{cc}
M_1 & 0 \\ 0 & M_2
\end{array}
\right)  \ . \label{DM2}
\end{equation}
Again here $M_1$ and $M_2$ ($M_1 \leq M_2$) are of the order of
the heavy Majorana masses, and $a, b, c, d \ \sim  \ M_j
|s_L^{\nu_l}|$. By diagonalization of matrix ${\cal M}$ [see
Eq.~(\ref{congruent}), but this time with $U$ orthogonal] we get
the neutrino masses
\begin{eqnarray}
m_1 &=& m_2 = 0 \ , \nonumber\\ m_3 &=& m_4 = M_1 \left[ 1 +
\frac{1}{2} \frac{a^2 + c^2}{M_1^2} + \cdots \right] \ ,
\nonumber\\ m_5 &=& m_6 = M_2 \left[ 1 + \frac{1}{2} \frac{b^2 +
d^2}{M_2^2} + \cdots \right] \ , \label{emassm2}
\end{eqnarray}
where the dots represent terms $\sim 1/M_j^4$. The phase factors
in Eq.~(\ref{congruent}) are $\lambda_3 = \lambda_5 = -1$,
$\lambda_1 = \lambda_2 = \lambda_4 = \lambda_6 = +1$ and the
mixing elements $B_{l h}$ are real, fulfill the symmetry $B_{l 3}
= B_{l 4}$ and $B_{l 5} = B_{l 6}$ (for $l=e, \mu$), and are given
explicitly by:
\begin{eqnarray}
B_{l h}  =  \frac{1}{\sqrt{2}}\left[
\begin{array}{llll}
 -a/M_1 \ , &  -a/M_1 \ , & b/M_2 \ ,  b/M_2 \\
 -c/M_1 \ , & -c/M_1 \ , & d/M_2 \ , d/M_2
\end{array}
\right] + {\cal O} \left( m_D^3/m_M^3\right)\ldots , \quad \ell =
e,\, \mu\, ;\ h= 3,\, 4,\, 5,\, 6\label{Blh2}
\end{eqnarray}

Notice that the two ``light'' neutrinos are actually massless,
although they can easily acquire small masses by introducing small
nonzero entries in the lower right $2 \times 2$ block of the mass
matrix ${\cal M}$. The two degenerate pairs of heavy Majorana
neutrinos can be interpreted as two heavy Dirac neutrinos. The
symmetry in the masses, mixings and phase factors imply that the
contribution to $G_{M\bar M}$  from the Majorana boxes of Fig.~2,
[{\it i.e.} the terms with $G_{\rm Box}(x_I,x_J)$ in
Eq.~\ref{FBoxred}], cancel out. This contrasts with Model I where
such terms are important, and it is related to the fact that Model
II conserves total lepton {\it number} while Model I does not.

As before, we search for the maximum of $|{G_{M\bar M}}/G_F|$.
Since the $G_{\rm Box}(x_I,x_J)$ terms cancel, and the mixing
elements $B_{l h}$ are real in absence of CP violation, we can
write
\begin{eqnarray}
 |{G_{M\bar M}}/G_F| &=&  \frac{\alpha_W}{32\pi}\times {\Big |} \sum_{I,J = \rm 3}^{6}
\Big\{ 2 B_{e I} B_{e J} B_{\mu I} B_{\mu J} \ \left[ F_{\rm
Box}(x_I, x_J) - 2 F_{\rm Box}(x_I,0) + F_{\rm Box}(0,0) \right]
{\Big |}
 \nonumber\\
 & \leq & \frac{\alpha_W}{32\pi} \hat F_{\rm Box} \left( \sum_{I=3}^6 B_{e
I} B_{\mu I} \right)^2 \ . \label{apprm2}
 \nonumber\\
 & \leq & \frac{\alpha_W}{32\pi}  \hat F_{\rm Box}
\; (s_L^{\nu_e})^2 (s_L^{\nu_{\mu}})^2 \ , \label{Schwm2}
\end{eqnarray}
where $\hat F_{\rm Box} = 2{\rm max}|F_{\rm Box}(x_I,x_J)- 2
F_{\rm Box}(x_I,0) +F_{\rm Box}(0,0)|$.  The last inequality is
Schwarz's inequality, which saturates when the following
proportionalities are reached:
\begin{equation}
\frac{B_{e3}}{B_{\mu 3}} = \frac{B_{e4}}{B_{\mu 4}} =
\frac{B_{e5}}{B_{\mu 5}} = \frac{B_{e6}}{B_{\mu 6}} \ .
\label{propor}
\end{equation}
The first and the third equalities are fulfilled automatically.
The second equality, however, can be shown to be fulfilled only
when $ad = bc$. Thus, the maximum of $|G_{M\bar M}|$ is reached
when $(s_L^{\nu_e})^2$ and $(s_L^{\nu_{\mu}})^2$ are at their
upper bounds, Eqs.~(\ref{s2s})
\begin{eqnarray}
(s_L^{\nu_e})^2 & = & \frac{a^2}{M_1^2} + \frac{b^2}{M_2^2}\  +\
{\cal O}(m_D^4/m_M^4)\  = \  0.005 , \label{snuemax}
 \\
(s_L^{\nu_\mu})^2 & = & \frac{c^2}{M_1^2} + \frac{d^2}{M_2^2}\  +\
{\cal O}(m_D^4/m_M^4)\  = \  0.002 , \label{snumumax}
\end{eqnarray}
and the condition $ad = bc$ is simultaneously fulfilled. This
implies, for given values of high mass parameters $M_1$ and $M_2$,
three relations for the four low mass parameters $a$, $b$ $c$ and
$d$. Thus, we still have one degree of freedom for fixing the
values of low mass parameters, without much effect on the
conversion probability. The simplest additional relation seems to
be the symmetry assumption $b = c$. All this then gives the
optimal choice of low mass parameters $a$, $b$, $c$ and $d$ in
Model II:
\begin{eqnarray}
a & = & \frac{M_2}{\sqrt{(M_2/M_1)^2 +
(s_L^{\nu_{\mu}}/s_L^{\nu_e})^2 }} \frac{s_L^{\nu_e}}{\sqrt{1 -
{s_L^{\nu_e}}^2 - {s_L^{\nu_{\mu}}}^2 }} \ , \label{aoptm2}
\\
b & = & c = a \times (s_L^{\nu_{\mu}}/s_L^{\nu_e}) \ , \qquad d =
a \times (s_L^{\nu_{\mu}}/s_L^{\nu_e})^2 \ , \label{bcdoptm2}
\end{eqnarray}
where $s_L^{\nu_e}$ and $s_L^{\nu_{\mu}}$ are taken at their
respective upper bound. Again, we have reduced the parameter space
to just two quantities, $M_1$ and $M_2$. Fig.~4 shows the graphs
for the maximal effective coupling $|G_{M\bar M}|$ as a function
of $M_1$ and $M_2$, which is consistent with all present
constraints within model II.

\begin{center}
\begin{figure}[htb]
\epsfig{file=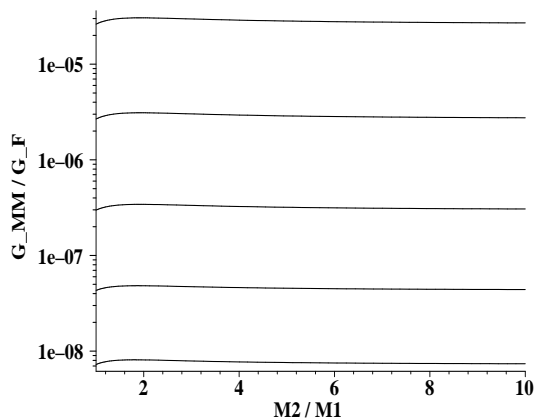,height=7.5 cm,width=6 cm,angle=-90}
\caption{ Maximal effective coupling $G_{M\bar M}/G_F$ for muonium
to antimuonium conversion in Model II, as a function of the ratio
of Majorana masses $M_2/M_1$, for different masses $M_1$: from the
lower to the upper curve, $M_1 =$ 100 GeV, 300 GeV, 1 TeV, 3 TeV,
10 TeV. }
 \label{fig4}
\end{figure}
\end{center}

We see that in this model the amplitude $G_{M\bar M}$ is an
increasing function of $M_1$, but is rather insensitive to $M_2$
 ($M_2>M_1$ by definition). Comparing Figs.~3
and 4 we see that in model II the maximum probability can reach
slightly larger values than in model I. However, there is a
stronger distinction between the models: here the maxima are
reached for natural values of the parameters, while the maxima in
model I are only reached for a finely tuned condition, which
circumvents the seesaw suppression of the mixing; away from this
finely tuned condition, the mixing suppression is very strong,
pushing the amplitude of model I several orders of magnitude below
the maximum.

\section{Prospects for Free $e^- \mu^+ \to \mu^- e^+$ Collisions}
\label{boxcs}

As an aside from this calculation, one can notice that the same
amplitude obtained from the box diagrams can be used to determine
the cross section for free $e^- \mu^+ \to \mu^- e^+$ collisions.
The result is simply:
\begin{eqnarray}
\sigma ( e^- \mu^+ \to \mu^- e^+) & = & \frac{4\pi}{3}
\frac{\alpha_w^2}{ M_W^4}  s {\big |} G_{M\bar M}/G_F {\big |}^2
\nonumber\\& \sim & 3\times 10^{-5} \left(\frac{s}{M_W^2}\right)
\left( \frac{|G_{M\bar M}|/G_F}{10^{-5}} \right)^2 \ {\rm fbarn}.
\label{sig}
\end{eqnarray}
 This expression is applicable for CMS
energies well above $m_{\mu}$ but below $M_W$, i.e., for 1 GeV $ <
\sqrt{s} < 10^2$ GeV. In this kinematical regime, the cross
section is clearly too small to be observed in the foreseeable
future, especially considering that both expressions in
parentheses are at most of order unity. For this reason, we will
comment on this process no further.

\section{Conclusions}
We have looked for the maximum probabilities of muonium to
antimuonium conversion that can be reached in models where the
required lepton flavor violation is caused by heavy neutrinos,
keeping consistency with all present bounds. The estimates are
done within two alternative models. In both models the amplitudes
are in general growing functions of the heavy neutrino masses. Our
estimates of the amplitudes are thus limited by two effects: (i)
suppression of flavor mixing matrix elements as the masses get
larger and (ii) the breakdown of perturbative expansion for masses
above a few tens of TeV.

Model I includes extra right handed neutrino singlets and
constitutes a typical seesaw model for the neutrino masses and
mixings, which violates lepton flavor as well as lepton number.
Model II includes extra right handed and left handed neutrino
singlets in such a way that lepton flavor is violated, yet lepton
number is conserved. We found that in Model I the Majorana boxes
give the dominant contribution over the Dirac boxes in the most
typical scenarios, yet the maximum probabilities are obtained in
finely tuned regions of the parameter space, where Dirac and
Majorana boxes are comparable and even tend to interfere
destructively. As a consequence of this effect, the maximal
conversion rates are not as large as one could have na{\"\i}vely
estimated. The maximal couplings thus obtained are $G_{M\bar
M}/G_F\sim 10^{-5}$, which means conversion probabilities 4 orders
of magnitude below present experimental sensitivities. Moreover,
away from the finely tuned condition, the small masses of the
light neutrinos enforce a dramatic suppression of the mixing (a
typical seesaw behavior) and the probabilities drop by several
orders of magnitude to unreachable levels.

Model II, in contrast, requires no finely tuned condition to reach
its maximal amplitudes. In this case, because of lepton number
conservation, the heavy Majorana neutrinos come in degenerate
pairs, thus acting effectively as Dirac neutrinos. As a result,
the Majorana boxes cancel in the amplitudes and the standard Dirac
boxes, which are generally smaller, are the only contribution.
Therefore, the cancellation of the Majorana boxes occurs only in
model II and it is due to interference between the different
neutrino flavors ({\it i.e.} the flavor mixing matrix), and not
because of Fermi statistics, as was stated in ref.\,\cite{Clark}.
Despite the cancellation of the Majorana boxes, Model II gives the
largest possible conversion probability, because the mixings are
free from the seesaw suppression of model I imposed by the light
neutrino masses. Overall, the effective coupling $G_{M\bar M}/G_F$
in model II can reach $10^{-5}\to 10^{-4}$ for natural values of
the parameters and neutrino masses around 10 TeV.

We also included the calculation of free $\mu^+e^-\to\mu^-e^+$
scattering at high energies, because this process tests the same
amplitude as muonium-antimuonium conversion. However the cross
section is so small that muonium-antimuonium conversion is still a
more hopeful alternative from the experimental point of view.

\acknowledgments
C.D.\ and G.C.\ are grateful to S. Kovalenko for
useful discussions. C.D. acknowledges support from Fondecyt
(Chile) research grant No. 1030254 and G.C. from Fondecyt grants
No.~1010094 and No.~7010094. The work of C.S.K. was supported in
part by the CHEP-SRC Program and by Grant No. R02-2003-000-10050-0
from BRP of the KOSEF (Korea). The work of J.D.K was supported by
the Korea Research Foundation Grant (2001-042-D00022).

\begin{appendix}
\section[]{The Models}
\setcounter{equation}{0}

In this Appendix we will briefly describe two distinct seesaw-type
models, which are used in our numerical evaluation of the
muonium-antimuonium conversion probability. Basically, these
models have extended neutrino sectors as compared with the SM. The
entire lepton sector has the same spectrum as the Standard Model
(we consider $N_L$ families in general) plus $N_H$ extra
right-handed neutrinos $\widetilde{\nu}_{R}^{(j)}$ ($j = 1 ,
\ldots, N_H$). The lepton sector is thus divided in $N_L$ ${\rm
SU(2)}_L$-doublets $L^{(i)}$, $N_L$ charged singlets $l_{R}^{(i)}$
and $N_H$ neutral singlets ${\widetilde \nu}_{R}^{(j)}$
\begin{eqnarray}
L^{(i)} & = & \left(
\begin{array}{c}
\nu_{ L}^{(i)} \\ l_{L}^{(i)}
\end{array}
\right) \ , \quad l_{R}^{(i)} \ , \quad {\widetilde \nu}_{R}^{(j)}
\ , \quad (i = 1, \ldots, N_L; \ j = 1,\ldots, N_H) \ .
\label{qL}
\end{eqnarray}
Concerning the interactions, only those involving charged currents
are relevant to our case (lepton currents coupled to $W$'s and, in
a general gauge, to charged Goldstone bosons $G^{\pm}$):
\begin{eqnarray}
{\cal L}_{l n W}(x) & = & \left( - \frac{g}{\sqrt{2}} \right)
\left[ {\bar l}(x) \gamma_{\mu} {\nu}_{L}(x) \right] W^-(x)^{\mu}
+ {\rm h.c.} \ ,
\label{lnWfl}
\\
{\cal L}_{l n G}(x) & = & \left( - \frac{g}{\sqrt{2} M_W} \right)
G^-(x) \left[ {\bar l}(x) m_l^{\dagger} \nu_L(x) - {\bar l}(x) m_D
{\widetilde \nu}_R(x) \right] + {\rm h.c.} \ .
\label{lnGfl}
\end{eqnarray}
In Eqs.~(\ref{lnWfl})-(\ref{lnGfl}), we used the following
simplified notation: $l$ is a column with the $N_L$ charged
leptons, $l_i = l_{L}^{(i)} + l_{R}^{(i)}$; $\nu_L$ is a column
with the $N_L$ standard neutrinos, $\nu_{L}^{(i)}$; ${\widetilde
\nu}_R$ is a column with the $N_H$ extra neutrinos ${\widetilde
\nu}_{R}^{(j)}$; $m_l$ is the standard $N_L \times N_L$ mass
matrix of the charged leptons (which can be taken to be diagonal
and positive, without loss of generality) ; $m_D$ is a new $N_L
\times N_H$ mass matrix of Dirac type for the neutrinos. $g$ is
the $SU(2)_L$ coupling constant ($g^2 = 8 G_F M_W^2/\sqrt{2}$).

Expression (\ref{lnGfl}) is obtained from the Yukawa terms for the
leptons and the Higgs doublet $\Phi$ (or its conjugate $\Phi^c =i
\tau_2 \Phi^{\ast}$):
\begin{eqnarray}
{\cal L}_{Y}(x) & = & \sum_{i,j=1}^{N_L} {\overline {L^{(i)}} }(x)
\,\Phi(x)\, (-{\cal D}_{i j}) l_R^{(j)}(x) + \sum_{j=1}^{N_H}
\sum_{i=1}^{N_L} { \overline {L^{(i)}} }(x)\, \Phi^c(x)\, ( -
{\cal U}_{i j}) {\widetilde \nu}_R^{(j)}(x) + {\rm h.c.}
\label{LY}
\end{eqnarray}
The first term is standard and gives the mass to the charged
leptons in terms of the Yukawa couplings ${\cal D}_{i j}$ when the
Higgs acquires a vev $\langle\Phi\rangle =(0, v)^T/\sqrt{2}$,
namely $[m_l]_{i j}={\cal D}_{i j}v/\sqrt{2}$; it also generates
the first interaction term in Eq.\ (\ref{lnGfl}). The second term
is extra physics; in a similar way it generates Dirac type masses
for the neutrinos, $[m_D]_{i j}={\cal U}_{i j} v/\sqrt{2}$ and the
second interaction term in Eq.\ (\ref{lnGfl}).

In addition, in these seesaw models there are neutrino mass terms
of Majorana type, $\propto {\overline {{\widetilde \nu}_R^{ \ c}}}
{\widetilde \nu}_R + {\rm h.c.}$, which can be gathered in a
matrix $m_M$ of dimension $N_H \times N_H$. Consequently, the
totality of neutrino mass terms can be expressed in the form:
\begin{equation}
-{\cal L}^{\nu}_{\rm mass}=
\frac{1}{2} \left( {\overline {{\nu}_L}},
{\overline {{\widetilde \nu}_R^{ \  c}}}   \right)
{\cal M}
\left(
\begin{array}{c}
\nu_L^{ \  c} \\ {\widetilde \nu}_R
\end{array} \right) +  {\rm h.c.}
= \frac{1}{2} \left( {\overline {{\nu}_L}},
{\overline {{\widetilde \nu}_R^{ \  c}}}   \right)
\left(
\begin{array}{cc}
0 & m_D \\ m^T_D & m_M
\end{array} \right)
\left(
\begin{array}{c}
\nu_L^{ \  c} \\ {\widetilde \nu}_R
\end{array} \right) +  {\rm h.c.} \ .
\label{Lnumass}
\end{equation}
The superscript ``$c$'' denotes charge-conjugated fields $\psi^c =
{\cal C} {\overline \psi}^T$, where ${\cal C} = - i \gamma^2
\gamma^0$ in the Dirac representation. The block in the upper left
corner is zero due to the $\rho$-parameter constraints (see e.g.\
Ref.~\cite{Kayserb}). The total mass matrix ${\cal M}$ appearing
in Eq.~(\ref{Lnumass}) has dimension
$(N_L\!+\!N_H)\!\times\!(N_L\!+\!N_H)$ and is a symmetric, in
general complex, matrix with a texture leading to a seesaw
mechanism upon diagonalization. This matrix can always be
diagonalized by means of a congruent transformation involving a
unitary matrix $U$ \cite{Aguilar-Saavedra:1996ev}
\begin{equation}
U{\cal M}U^{T} \Lambda^* = {\cal M}_d \ .
\label{congruent}
\end{equation}
Here, ${\cal M}_d$ is a nonnegative diagonal matrix, and
$\Lambda^*$ is a diagonal unitary matrix: $(\Lambda^*)_{ij} =
\delta_{ij} \lambda^*_i$, where $\lambda_i$ are complex phase
factors ($|\lambda_i|=1$) such that ${\cal M}_d$ is made real and
nonnegative. The diagonal $\Lambda^*$ matrix in
Eq.~(\ref{congruent}) can be incorporated into $U$ by the
redefinition: $U_{\rm new} = (\Lambda^*)^{1/2} U$, where $U_{\rm
new}$ is again unitary; $\Lambda^*$ thus reflects a freedom of
choice for $U$. The $N_L\!+\!N_H$ mass eigenstates $n_i$ are
Majorana neutrinos, related to the interaction eigenstates $\nu_a$
by the matrix $U$ of Eq.~(\ref{congruent})
\begin{eqnarray}
\left(\begin{array}{c} \nu_L \\
{\widetilde \nu}_R^{ \  c} \end{array} \right)_a
&=&
\sum_{i=1}^{N_L+N_H} U^{*}_{ia} ~{n_{i L}}
\quad
\Rightarrow \
\quad
\left(\begin{array}{c} \nu_L^{ \  c} \\
{\widetilde \nu}_R \end{array} \right)_a
=
\sum_{i=1}^{N_L+N_H}
U_{ia} ~\lambda^{*}_{i}~{n_{i R}} \ ,
\label{nuvsn}
\end{eqnarray}
The first $N_L$ mass eigenstates are the light standard partners
of the charged leptons, while the other $N_H$ eigenstates are
heavy (if the seesaw mechanism takes place). The factor
$(\Lambda^*)_{ii} = \lambda^{*}_i$ is now recognized as the
creation phase factor \cite{Kayserb,Kayser:1984ge} of the Majorana
neutrino $n_i$ $(\equiv n_{i L}\!+\!n_{i R})$, in the sense that
$n_i^c = \lambda^*_i n_i$, and more specifically $(n_{i L})^c =
\lambda_i^* n_{i R}$. The
$N_L\!\times\!(N_L\!+\!N_H)$--dimensional mixing matrix $B$ for
charged current interactions is defined as
\begin{equation}
B_{l i}=U^{*}_{i l}  = \left( U^{\dagger} \right)_{l i} \ .
\label{B}
\end{equation}
When CP is conserved, the coefficients $B_{l i}$ can be chosen
real and all the phase factors $\lambda_i$ real as well
\cite{Kayserb}: $\lambda_i = {\tilde \eta}_{\rm CP}(n_i)/i = \pm
1$, where ${\tilde \eta}_{\rm CP}(n_i)=\pm i$ is the intrinsic CP
parity of the Majorana neutrino $n_i$. If the symmetric mass
matrix ${\cal M}$ is real, then there exists a unitary real (i.e.,
orthogonal) matrix $U={\cal O}$ such that
\begin{equation}
{\cal O} {\cal M} {\cal O}^T = {\cal D} \ ,
\label{orth}
\end{equation}
where ${\cal D}$ is a diagonal and real matrix; in general, some
of its diagonal elements are negative, but the absolute values are
the same as those of ${\cal M}_d = {\cal D} \Lambda^*$: $({\cal
M}_d)_{ii} = {\cal D}_{ii} \lambda^*_i = |{\cal D}_{ii}|$. We see
that $\lambda_i$ is here $\pm 1$ ($= {\rm sgn}[{\cal D}_{ii}]$).
Therefore, when the symmetric mass matrix ${\cal M}$ is real, we
are in a situation with no CP violation: $B_{l i} = {\cal O}_{i
l}$ are real, and $\lambda_i = \pm 1$. Even more, in that case
${\cal O}_{ij}$ for $j > N_L$ are also real. Thus, the reality of
${\cal M}$ is sufficient for having CP conservation.\footnote{It
is not clear to us what are the necessary and sufficient
conditions on the elements of ${\cal M}$ for having CP
conservation. If only the upper left $N_L \times N_L$ block of
${\cal M}$ is real, it appears that we can have CP violation,
i.e., the elements $B_{l i} = U^*_{i l}$ ($l = 1, \ldots, N_L$)
apparently cannot always be chosen real in such a case.} We will
see later that in this case the values of $\lambda_i$ either as
$+1$ or $-1$, separately for each $i$ ($i=1,\ldots,N_L+N_H$),
influence our results.

Inserting the transformations (\ref{nuvsn}) in the interaction
terms (\ref{lnWfl}) and (\ref{lnGfl}), taking also into account
the mass matrix form (\ref{Lnumass}) and the transformation
(\ref{congruent}) into Eq.\ (\ref{lnGfl}), we obtain the explicit
interactions of the charged bosons with the leptons in their mass
basis:
\begin{eqnarray}
{\cal L}_{l n W}(x) & = & \left( - \frac{g}{\sqrt{2}} \right)
\sum_{i=1}^{N_L} \sum_{j=1}^{N_L+N_H} \left[ B_{ij} {\bar l}_i(x)
\gamma_{\mu} P_L n_j(x) \right] W^-(x)^{\mu} + {\rm h.c.} \ ,
\label{lnW}
\\
{\cal L}_{l n G}(x) & = & \left( - \frac{g}{\sqrt{2} M_W} \right)
 G^-(x) \sum_{i=1}^{N_L} \sum_{j=1}^{N_L+N_H}
\left[ B_{ij} {\bar l}_i(x) \left(m_i P_L - M_j P_R \right)
n_j(x) \right] + {\rm h.c.}  \ ,
\label{lnG}
\end{eqnarray}
where $m_i > 0$ is the mass of the charged lepton $l_i$, and $M_j
\geq 0$ the mass of the Majorana neutrino $n_j$. Expressions
(\ref{lnW}) and (\ref{lnG}) were obtained already in
Ref.~\cite{Pilaftsis:1991ug}, and used in Refs.~\cite{IP,CDKK} and
others.

In this work, we have used two specific models:

\medskip
{\bf Model I:} \ \ This is the usual seesaw model described above,
with $N_L$ left-handed neutrinos $\nu_{i L}$ and an equal number
of right-handed neutrinos ${\widetilde \nu}_{i R}$: $N_H=N_L$. The
neutrino mass terms are those given in Eq.~(\ref{Lnumass}). The
square ($N_L\!\times\!N_L$) matrices $m_M$ and $m_D$ are supposed
to be invertible.

The heavy-to-light neutrino mixings
$(s_L^{\nu_l})^2 \equiv \sum_{h} | U_{h l} |^2$
are of the order of the  squared ratio between the Dirac mass ($m_D$)
and the Majorana mass ($m_M$) scales ($\sim\!|m_D|^2/|m_M|^2$).
The eigenmasses of the light neutrinos are
$m_{\nu_{light}}\!\sim\!m^2_D/m_M$.
Severe constraints on the
$|m_D| \ll |m_M|$ hierarchy in Model I are
imposed by the
very low experimental bounds
$m_{\nu_{light}} \stackrel{<}{\sim} 1$ eV.
Another set of constraints on the model
is imposed by the present experimental
bounds on the heavy-to-light mixing parameters $(s_L^{\nu_l})^2
\sim |m_D|^2/|m_M|^2$ ($\stackrel{<}{\sim} 10^{-2}$, see below).

\medskip
\noindent{\bf Model II:}\ \  This model is formally again of the
seesaw type, because the mass matrix ${\cal M}$ has a form similar
to (\ref{Lnumass}). However, now $N_H=2 N_L$. This model contains
an equal number $N_L$ of left-handed ($S_{i L}$) and right-handed
(${\widetilde \nu}_{i R}$) neutral singlets~\cite{E6,SO10a}, and
thus the right-handed column ${\widetilde \nu}_R$ with $N_H$
components is now (${\widetilde \nu}_{R}, S_{L}^{ \ c})^T$. The
form of the mass matrix ${\cal M}$ in this model ensures
conservation of total lepton number, but lepton flavor mixing is
still possible. The neutrino mass terms, after electroweak
symmetry breaking, have the form
\begin{equation}
-{\cal L}^{\nu}_{\rm mass}=\frac{1}{2}
\left( {\overline {{\nu}_L}},{\overline {{\widetilde \nu}_R^{ \ c}}},
{\overline {{S}_L}} \right)
{\cal M}
\left(
\begin{array}{c}
{\nu_L^{ \ c}} \\ {\widetilde \nu}_R \\ S_L^{ \ c}
\end{array}
\right)
+ {\rm h.c.} \ , \qquad
{\cal M}= \left(
\begin{array}{ccc}
0 & m_D & 0\\ m_D^T & 0 & m_M^T \\ 0 & m_M & 0
\end{array} \right) \ ,
\label{M-model2}
\end{equation}
which is formally of the form (\ref{Lnumass}), but with a specific
$N_L\!\times\!N_L$ texture: the Dirac mass submatrix now has a
specific $N_L\!\times\! 2 N_L$ form $(m_D,0)$, where $m_D$ is a
square ($N_L\!\times\!N_L$) matrix, and $0$ is the
$N_L\!\times\!N_L$ zero matrix; and the Majorana mass submatrix
$m_M$ of (\ref{Lnumass}) is now replaced by a $2N_L\!\times\!2N_L$
matrix in the lower right part of ${\cal M}$ (\ref{M-model2}). As
given in (\ref{M-model2}), the matrix ${\cal M}$ gives for each of
the $N_L$ generations a massless Weyl neutrino and two degenerate
Majorana neutrinos \cite{BRV,Gonzalez-Garcia:1989rw}.
Consequently, the seesaw-type restriction
$m_{\nu_{light}}\!\sim\!m^2_D/m_M \stackrel{<}{\sim} 1$ eV of
Model I is absent in Model II in its unperturbed form
(\ref{M-model2}). Nevertheless, the experimental bounds on the
heavy-to-light mixing parameters $(s_L^{\nu_l})^2 \sim
|m_D|^2/|m_M|^2$ ($\stackrel{<}{\sim} 10^{-2}$) do impose a
hierarchy $|m_D| < |m_M|$ between the Dirac and Majorana mass
sectors and, in this sense, the model remains of a seesaw type,
but the hierarchy $|m_D| < |m_M|$ here is in general much weaker
than in Model I. Interestingly, nonzero masses for the $N_L$ light
neutrinos can easily be included in Model II by introducing small
Majorana mass terms for the neutral singlets $S_{i L}$, i.e.,
small nonzero elements in the $N_L\!\times\!N_L$ lower right block
of ${\cal M}$. The mixings of heavy-to-light neutrinos are not
significantly affected by these small corrections.

The above two models, for specific choices of the
$N_L\!\times\!N_L$ matrices $m_D$ and $m_M$, give us specific
values of the mixing-matrix elements (\ref{B}) and eigenmasses
$M_j$, i.e., the parameters which then determine the charge
current interactions of (\ref{lnW}) and (\ref{lnG}).

\section[]{Box diagram amplitudes with Majorana neutrinos}
\setcounter{equation}{0}

Here we outline the derivation of the expression for $G_{M\bar M}$
in Eq.\ (\ref{FBoxeuue}) at one loop order (box diagrams), defined
by the effective lagrangian of Eq.\ (\ref{Leff}).

There are basically four types of box diagrams: two containing
generic neutrinos (Fig.~1) and two exclusive for Majorana
neutrinos only (Fig.~2). According to the contractions of the
fields with the external states, the diagrams of Fig.~1.b, 2.a and
2.b have the same relative sign, while the diagram of Fig.~1.a has
the opposite sign.

There are also basically two momentum integrals encountered in the
box diagrams, in the limit where external masses and momenta are
neglected compared to the loop momenta. The first one is:

\begin{eqnarray}
I_{i j} & \equiv & \int \frac{d^4 q}{ (2 \pi)^4 } \frac{1}{ (q^2\!
-\! M_i^2) (q^2\! -\! M_j^2) (q^2\! -\! M_W^2)^2 } = \frac{i}{ (4
\pi)^2 M_W^4 } {\cal J}( x_i, x_j )  ,
\label{Iij0}
\end{eqnarray}
where we denote $x_j = M_j^2/M_W^2$ and where ${\cal J}(x_i,x_j)$
is a dimensionless expression:
\begin{eqnarray}
{\cal J}(x_i,x_j ) & = &
- \frac{1}{( x_i - x_j )} \left\{  \left( \frac{1}{ (1 - x_i
)}+ \frac{ x_i \ln x_i } { (1 - x_i )^2 }  \right)\  -\
\left(x_i\to x_j \right)\ \right\}  .
\label{J2}
\end{eqnarray}
The second integral is:
\begin{eqnarray}
K_{i j} & \equiv & \int \frac{d^4 q}{ (2 \pi)^4 } \frac{q^2}{
(q^2\! -\! M_i^2) (q^2\! - \!M_j^2) (q^2\! -\! M_W^2)^2 } =
\frac{i}{ (4 \pi)^2 M_W^2 }  {\cal K}( x_i,x_j )  , \label{Kij0}
\end{eqnarray}
where the dimensionless expression is:
\begin{eqnarray}
{\cal K}(x_i,x_j ) & = &
- \frac{1}{( x_i - x_j )}
 \left\{ \left(\frac{1}{1-x_i} + \frac{ x_i^2 \ln x_i }
{ (1 - x_i )^2 }\right)  - \ \left(x_i\to x_j\right) \right\}  .
\label{K2}
\end{eqnarray}

The transition amplitudes ${\cal T}^{(1a)}_{WW}$ and ${\cal
T}^{(1b)}_{WW}$, which correspond to the diagrams of Figs.~1.a and
1.b when both internal bosons are $W$, are:
\begin{eqnarray}
\left\{
\begin{array}{c}
{\cal T}^{(1a)}_{WW}
\\
{\cal T}^{(1b)}_{WW}
\end{array}
\right\} & = & \left( - \frac{g}{ \sqrt{2} } \right)^4
B_{e i}^{\ast} B_{e j}^{\ast} B_{\mu i} B_{\mu j}
\nonumber\\
&&
\times \int \frac{d^4 q}{ (2 \pi)^4 }
\frac{1}{ (q^2\!-\! M_i^2) (q^2\! -\! M_j^2) (q^2\! -\! M_W^2)^2 }
\left\{
\begin{array}{r}
(+1) \left[ {\overline u}(k^\prime) \gamma_{\mu} {q \llap / }
\gamma_{\nu} P_L v(p^\prime) \right] \left[ {\overline v}(k)
\gamma^{\nu} {q \llap / } \gamma^{\mu} P_L u(p) \right] \\ (-1)
\left[ {\overline u}(k^\prime) \gamma_{\mu} {q \llap / }
\gamma_{\nu} P_L u(p) \right] \left[ {\overline v}(k) \gamma^{\nu}
{q \llap / } \gamma^{\mu} P_L v(p^{\prime}) \right]
\end{array}
\right\}   . \label{TWWab1}
\end{eqnarray}
Overall we label $p$ and $p^\prime$ as the momenta of the initial
and final electron (positron), and $k$ and $k^\prime$ as  the
initial and final momenta of the (anti-) muons. To simplify the
notation, we have omitted the spin and flavor labels in the
spinors, since the momenta specify them unambiguously,
(c.f.~Fig.~1). We notice a factor $(-1)$ in the second amplitude;
its origin is the anticommutation of the fermion fields involved
in the Wick contractions with the external states. Now, using the
identity
\begin{eqnarray}
\gamma_{\mu} \gamma_{\alpha} \gamma_{\nu}= i \epsilon_{\lambda \mu
\alpha \nu} \gamma^{\lambda} \gamma_5 + \gamma_{\mu} g_{\alpha
\nu}+ \gamma_{\nu} g_{\mu \alpha} - \gamma_{\alpha} g_{\mu \nu}
\label{gggid}
\end{eqnarray}
the spinor structure can be simplified to the following form (we
have omitted the spinors in the brackets):
\begin{equation}
\left[  \gamma_{\mu} {q \llap / } \gamma_{\nu} P_L \right] \left[
\gamma^{\nu} {q \llap / } \gamma^{\mu} P_L \right] = 4 q^{\alpha}
q^{\beta} \left[ \gamma_{\alpha} P_L \right] \left[ \gamma_{\beta}
P_L \right]  .\label{idqslash}
\end{equation}
Finally, after replacing $q^{\alpha} q^{\beta} \to g^{\alpha
\beta} q^2/4$ due to the Lorentz invariance of the integrals and
doing the following Fierz transformation in the second amplitude:
\begin{eqnarray}
\left[ {\overline u}_1 \gamma_{\alpha} P_L u_2 \right] \left[
{\overline v}_3 \gamma^{\alpha} P_L v_4 \right]= - \left[
{\overline u}_1 \gamma_{\alpha} P_L v_4 \right] \left[ {\overline
v}_3 \gamma^{\alpha} P_L u_2 \right] \label{Fierz1}
\end{eqnarray}
both amplitudes in Eq.~(\ref{TWWab1}) become identical:
\begin{eqnarray}
{\cal T}^{(1a)}_{WW} = {\cal T}^{(1b)}_{WW}
= i \frac{ \alpha_w^2}{ 4 M_W^2}
B_{e i}^{\ast} B_{e j}^{\ast} B_{\mu i} B_{\mu j}
{\cal K}( x_i,x_j ) \left[ {\overline u}(k')
\gamma_{\alpha} P_L v(p') \right] \left[ {\overline v}(k)
\gamma^{\alpha} P_L u(p) \right]  . \label{TWWab2}
\end{eqnarray}

In the gauge we are using, one must also include the diagrams with
one and two Goldstone bosons, {\it i.e.} ${\cal T}^{(1a)}_{WG}$,
${\cal T}^{(1a)}_{GW}$, ${\cal T}^{(1a)}_{GG}$, etc. The
calculation goes through similar steps as those already stated for
the box with two $W$'s, but contains different propagators and
therefore different combinations of the integrals (\ref{Iij0}) and
(\ref{Kij0}).

The sum of all these diagrams
 gives the total amplitude of the ``Dirac'' box diagrams of Fig.~1:
\begin{eqnarray}
{\cal T}^{(1)}  & \equiv & \sum_{\kappa = a,b} \left[ {\cal
T}^{(1\kappa)}_{WW} + {\cal T}^{(1\kappa)}_{GG} + {\cal
T}^{(1\kappa)}_{GW+WG} \right] \nonumber\\ & = & - i \frac{
\alpha_w^2}{4 M_W^2}\  2 B_{e i}^{\ast} B_{e j}^{\ast}  B_{\mu i}
B_{\mu j} \ F_{\rm Box}( x_i, x_j) \left[ {\overline u}(k^\prime)
\gamma_{\alpha} P_L v(p^\prime) \right] \left[ {\overline v}(k)
\gamma^{\alpha} P_L u(p) \right] , \label{Tab}
\end{eqnarray}
where $F_{\rm Box}( x_i, x_j)$ is given in Eq.\ (\ref{fbox}) and
is identical to the expression $F_{\rm Box}(x_i, x_j)$ of
Ref.~\cite{IP}.

The diagrams of Fig.~2 are, on the other hand, different in the
sense that they exist only if the internal neutrinos are of
Majorana type. The Wick contractions of the Majorana fields can be
performed in the usual way if we previously transform two of the
four vertex operators as follows: (we chose the vertices with the
initial anti-muon and with the final anti-electron):
\begin{subequations}
\label{Ctr2}
\begin{eqnarray}
{\overline \psi_\mu}(x) \gamma_{\nu} P_L n_j(x) & = & - \lambda_j
{\overline n_j}(x) \gamma_{\nu} P_R \psi_\mu^{c}(x)  ,
\label{Ctr2a}
\\
{\overline n}_i(x) \gamma_{\mu} P_L \psi_e(x) & = & -
\lambda_i^{\ast} {\overline {\psi_e^c}}(x) \gamma_{\mu} P_R n_i(x)
 . \label{Ctr2b}
\end{eqnarray}
\end{subequations}
Here, $\lambda_j$ is the creation phase factor of the Majorana
neutrino $n_j$, $i.e.$ $n_j^c = \lambda_j^{\ast} n_j$, and the
superscript ``$c$'' denotes the conjugated field $\psi^c = {\cal
C} {\overline \psi}^T$, where ${\cal C} = - i \gamma^2 \gamma^0$
in the Dirac representation. After performing the transformations
(\ref{Ctr2}), we can contract the ${\overline n}(x_k)$'s and
$n(x_n)$'s into normal neutrino propagators (see
Ref.~\cite{Kayserb}). Let us first consider the two diagrams of
Fig.~2 with two internal $W$'s. These diagrams differ from each
other only in the way the $W$'s are contracted. The amplitudes
are:
\begin{eqnarray}
\left\{
\begin{array}{c}
{\cal T}^{(2a)}_{WW}
\\
{\cal T}^{(2b)}_{WW}
\end{array}
\right\} & = & \left( - \frac{g}{ \sqrt{2} } \right)^4 \left( B_{e
i}^{\ast}\right)^2 \left( B_{\mu j}\right)^2 \lambda_j
\lambda_i^{\ast} \ M_j M_i (-1) \nonumber\\ && \times \int
\frac{d^4 q}{ (2 \pi)^4 } \frac{1}{ (q^2\!-\! M_i^2) (q^2\! -\!
M_j^2) (q^2\! -\! M_W^2)^2 } \left\{
\begin{array}{r}
\left[ {\overline u}(k^\prime) \gamma_{\mu} \gamma_{\nu} P_R u_c
(k) \right] \left[ {\overline u_c}(p^\prime) \gamma^{\mu}
\gamma^{\nu} P_L u(p) \right] \\ \left[ {\overline u}(k^\prime)
\gamma_{\mu} \gamma_{\nu} P_R u_c(k) \right] \left[ {\overline
u_c}(p^\prime) \gamma^{\nu} \gamma^{\mu} P_L u(p) \right]
\end{array}
\right\}  \ .
\label{TWWcd1}
\end{eqnarray}
We denoted as $u_c$ the $u$-spinors associated to the positively
charged fermions (usually considered ``antiparticles''), but
operationally these are normal $u$-spinors. The factor $(-1)$
before the integral is important and comes from the Wick
contractions of the fermion fields with the external states, given
the sign convention of the previous diagrams. Using the identities
\begin{equation}
\gamma^{\mu} \gamma^{\nu} = g^{\mu \nu} - i \sigma^{\mu \nu} \ ,
\qquad \sigma_{\mu \nu} \gamma_5 = \frac{i}{2} \epsilon_{\mu \nu
\alpha\beta} \sigma^{\alpha\beta} \ , \label{sigG5}
\end{equation}
the spinor structure in both amplitudes of Eq.~(\ref{TWWcd1}) can
be simplified to the same form:
\begin{equation}
\left[ \gamma_{\mu} \gamma_{\nu} P_R \right] \left[ \gamma^{\mu}
\gamma^{\nu} P_L \right]
=
\left[ \gamma_{\mu} \gamma_{\nu} P_R \right] \left[ \gamma^{\nu}
\gamma^{\mu} P_L \right] = 4 \left[ P_R \right] \left[ P_L \right]
. \label{WWcdid}
\end{equation}
This is the reason why the diagrams in Fig.~2 do not cancel but,
on the contrary, they add to twice their value.

Using the following Fierz identity:
\begin{eqnarray}
2\left[{\overline u}_1 P_R u_2 \right] \left[ {\overline u}_3 P_L
u_4 \right] = \left[{\overline u}_1 \gamma_{\alpha} P_L u_4
\right] \left[ {\overline u}_3 \gamma^{\alpha} P_R u_2 \right]
\label{Fierz2}
\end{eqnarray}
we can transform the spinor product into a product of currents.
The amplitudes of Eq.~(\ref{TWWcd1}) then result in:
\begin{eqnarray}
{\cal T}^{(2a)}_{WW} = {\cal T}^{(2b)}_{WW} & = & - i \frac{
\alpha_w^2}{ 2 M_W^2} \left( B_{e i}^{\ast}\right)^2 \left( B_{\mu
j}\right)^2 \lambda_j \lambda_i^{\ast} \ \sqrt{ x_i x_j } {\cal J}
(x_i, x_j) \left[{\overline u}(k^\prime) \gamma_{\alpha} P_L u(p)
\right] \left[ {\overline u_c}(p') \gamma^{\alpha} P_R u_c(k)
\right] \ . \label{TWWcd2}
\end{eqnarray}
Now, the diagrams with Goldstone bosons instead of $W$'s contain
scalar currents which, in analogy to Eqs.\ (\ref{Ctr2}), should be
transformed before doing the Wick contractions according to:
\begin{subequations}
\label{Ctr2s}
\begin{eqnarray}
{\overline \psi_\mu}(x) P_{L,R}\ n_j(x) & = &  \lambda_j
{\overline n_j}(x)  P_{L,R}\ \psi_\mu^{c}(x)  , \label{Ctr2as}
\\
{\overline n}_i(x) P_{L,R}\ \psi_e(x) & = &  \lambda_i^{\ast}
{\overline {\psi_e^c}}(x) P_{L,R}\ n_i(x) . \label{Ctr2bs}
\end{eqnarray}
\end{subequations}

Besides this detail, the calculation of the diagrams of Fig.~2
with one or two Goldstone bosons replacing the $W$'s follows the
same steps as above.

From Eq.~(\ref{TWWcd2}) and the analogous expressions
corresponding to the diagrams with Goldstone bosons,
it is clear that both diagrams of Fig.~2 give identical results.

We should now transform the $u$-spinors of the positively charged
fermions into $v$-spinors, as it is customary, using the following
conjugated current identity:
\begin{eqnarray}
{\overline u_c}(p') \gamma^{\alpha} P_R u_c(k) =  {\overline v}(k)
\gamma^{\alpha} P_L v(p')
\end{eqnarray}
and use the Fierz identity (\ref{Fierz1}) to arrive at the same
spinor structure as in the amplitude of Fig.~1, namely
Eq.~(\ref{Tab}). The total result for Fig.~2 is then:
\begin{eqnarray}
{\cal T}^{(2)} & \equiv & \sum_{\kappa = a,b} \left[ {\cal
T}^{(\kappa)}_{WW} + {\cal T}^{(\kappa)}_{GG} + {\cal
T}^{(\kappa)}_{GW+WG} \right] \nonumber\\ & = & i \frac{
\alpha_w^2}{ 4 M_W^2}\left( B_{e i}^{\ast}\right)^2 \left( B_{\mu
j}\right)^2 \lambda_j \lambda_i^{\ast} \ G_{\rm Box}(x_i, x_j)
\left[{\overline u}(k') \gamma_{\alpha} P_L u(p) \right] \left[
{\overline v}(k) \gamma^{\alpha} P_L v(p') \right] \ , \label{Tcd}
\end{eqnarray}
where the function $G_{\rm Box}$ is given in Eq.\ (\ref{gbox}),
which is identical to that of Ref.~\cite{IP}.

The sum of all diagrams is ${\cal T}^{(1)}+{\cal T}^{(2)}$ [
Eqs.~(\ref{Tab}) and (\ref{Tcd})]; from this amplitude one deduces
the efective coupling $G_{M\bar M}$ of Eq.~(\ref{FBoxeuue}).

\end{appendix}


\begin{thebibliography}{99}
\bibitem{spinor}
We use the standard relativistic norm for the states
$\{a_p,a_{p'}^\dagger\}=2E_p (2\pi)^3 \delta^3(p-p')$ and for the
spinors $u(p)^\dagger u(p) = 2 E_p$.
\bibitem{Fein-Wein}
Our mass mixing $m_{M\bar M}$ corresponds to $2\delta$
of Ref.\ \cite{Feinberg}.
\end{thebibliography}
\end{document}